\documentstyle[eqsecnum,aps,prd]{revtex}
\begin{document}
%======================================%
%<<<<<<<<<<<  DEFINITION  >>>>>>>>>>>>>%
%======================================%
%======================================%
%<<<<<<<<<<<< TITLE PAGE >>>>>>>>>>>>>>%
%======================================%
\thispagestyle{empty}
{\baselineskip0pt
\leftline{\large\baselineskip16pt\sl\vbox to0pt{\hbox{DAMTP} 
               \hbox{University of Cambridge}\vss}}
\rightline{\large\baselineskip16pt\rm\vbox to20pt{\hbox{DAMTP-99-100}
               \hbox{UTAP-354}
               \hbox{RESCEU-99-46}           
               \hbox{\today}
\vss}}%
}
\vskip20mm
\begin{center}
{\large\bf 
A Probe Particle in Kerr-Newman-deSitter Cosmos \\
}
\end{center}

\begin{center}
{\large Tetsuya Shiromizu}
\vskip 3mm
\sl{DAMTP, University of Cambridge \\ 
Silver Street, Cambridge CB3 9EW, United Kingdom\\
\vskip 5mm
Department of Physics, The University of Tokyo, Tokyo 113-0033, 
Japan \\
and \\
Research Centre for the Early Universe(RESCEU), \\ 
The University of Tokyo, Tokyo 113-0033, Japan
}
\vskip 5mm
{\large Uchida Gen}
\vskip 3mm
\sl{  Department of Earth and Space Science, Graduate School of Science,\\
Osaka University, Toyonaka 560-0043, Japan \\
and \\
Department of Physics, The University of Tokyo, Tokyo 113-0033, 
Japan}
\end{center}

%\begin{center}
%{\it }
%\end{center}
%======================================%
%<<<<<<<<<<<<< ABSTRACT >>>>>>>>>>>>>>>% 
%======================================%
\begin{abstract} 
We consider the force acting on a spinning charged test particle 
(probe particle) with the mass $m$ and the charge $q$ in slow rotating
the Kerr-Newman-deSitter(KNdS) black hole with the mass $M$ and the 
charge $Q$. We consider the case which 
the spin vector of the probe particle is 
parallel to the angular momentum vector of the KNdS space-time. 
We take account of the gravitational 
spin-spin interaction under the slow rotating limit of the KNdS 
space-time.  When $Q=M$ and $q=m$, we show that  
the force balance holds including the spin-spin interaction 
and the motion is approximately same as that of a particle in 
the deSitter space-time. This force cancellation suggests the
possibility of the existence of an exact solution of spinning 
multi-KNdS black hole. 
\end{abstract}
%\vfill
\vskip1cm

%\multicols{2}

%======================================%
%<<<<<<<<<<<< SECTION I  >>>>>>>>>>>>>>%
%======================================%
%\baselineskip25pt
\section{Introduction}

A half of decade ago the multi-black hole solution 
with a positive cosmological
constant has been discovered by Kastor and Traschen\cite{KT}. The 
Kastor-Traschen(KT) 
solution describes the dynamical system of multi-black holes. 
The solution is important because it offers a test of the cosmic 
censorship conjecture\cite{Brill}, a picture of two or several 
black holes collision\cite{Nakao} and an example supporting the cosmic no-hair 
conjecture\cite{nohair}. The issue on  
the differentiability of the cosmological horizon is also discussed
\cite{Bad}.  

In this paper we discuss the possibility of the
extension into the spinning version of the KT solution. 
In the limit of the zero cosmological constant, the KT solution becomes 
the Majumdar-Papapetrou(MP) solution\cite{MP}. 
The existence is closely related to  
the  static force cancellation as $Q=M$. 
The force balance can be confirmed by a charged probe 
particle. The spinning version of the MP solution has been constructed
by Israel-Wilson\cite{IW} and Perjes\cite{Perjes}. Unfortunately, 
the solution has naked singularities\cite{HH} when $Q=M$. 
On the other hand, in the Kerr-Newman-deSitter(KNdS) space-time, 
there are cases with no singularities 
even if $Q=M$. This implies that the exact solution is a 
spinning multi-{\it black-hole} solution if there is. 
This is a great advantage in the physical aspect. It is also important 
to consider an exact solution in the heterotic string theory on 
a torus here\cite{Sen1}. 
In higher dimensions than five, the multi-soliton solution has no 
singularities and the extreme limit is surprisingly identical 
with the saturation of the Bogomol'nyi bound\cite{Sen2}. 

In the spinning case, the force cancellation between a spinning particle
and the Kerr-Newman soliton with $Q=M$ has been confirmed up to and including
the gravitational spin-spin interaction\cite{KT2} by using a spinning charged 
probe particle. The equation of motion was 
derived in \cite{Papa}\cite{Dixon}\cite{Wald}.  
Hence it is natural to think that the force 
cancellation including the spin-spin interaction is closely related to 
the existence of the spinning multi-soliton solution,  
the IWP exact solution. Moreover, there is another explicit example of 
force cancellation in the Einstein-Maxwell-dilaton-axion(EMDA) system which
describes a low energy string theory\cite{Tess1}. The exact spinning 
multi-soliton solution 
definitely exists in the EMDA system\cite{EMDA}. On the other hand, the 
force cancellation is not guaranteed in the Einstein-Maxwell-dilaton(EMD)
system\cite{Tess2} for a probe particle having the same gyromagnetic
ratio as that of the background space-time. Although we cannot
definitely say, this situation might reflect the fact that the 
exact spinning 
multi-soliton solution has not been discovered yet in the EMD system. 

In this paper we will show the force cancellation between a spinning test 
particle and the KNdS space-time in order to search the possibility of the 
exact spinning multi-black hole solution in asymptotically deSitter 
space-times. 

The rest of this paper is organised as follows. In Sec. II, we review 
the equation of motion for a spinning charged particle following 
Dixon\cite{Dixon}. We also derive some features of the probe
particle which will be used later. In Sec. III, we calculate the 
force acting on the probe particle in the slow rotating KNdS 
space-time. For simplicity we consider the probe particle whose 
spin vector is parallel to the angular momentum vector of the 
KNdS space-time. 
We will confirm that the force cancellation holds including the 
gravitational spin-spin interaction as $Q=M$ and 
$q=m$ and show that the probe particle freely moves regardless of 
the central black hole. In Sec. IV, 
we will summarise our present study and discuss remaining problems. 
Some exact expressions of the basic equation are 
written down in the appendix A. In the appendix B, we give the energy
bound argument which the saturation yields $Q=M$. 

\section{Equation of Motion}

The equation of motion for a spinning charged particle was derived by 
Dixon\cite{Dixon} based on the Papapetrou's work\cite{Papa}. The basic
equations are\footnote{One can see the similar derivation in the 
refs\cite{spin1}.} 
%========<Equation>========%
%
\begin{eqnarray}
v^\nu \nabla_\nu p^\mu =-
\frac{1}{2}R^\mu_{~\nu \alpha\beta}v^\nu S^{\alpha\beta}
+qF^\mu_{~\nu}v^\nu+\frac{gq}{4m}\Bigr( \frac{p\cdot v}{-m}\Bigr) 
\nabla^\mu F_{\alpha\beta}S^{\alpha\beta} \label{eq:basic1}
\end{eqnarray}
%
%==========================%
and
%========<Equation>========%
%
\begin{eqnarray}
v^\mu \nabla_\mu S^{\alpha\beta}=
p^\alpha v^\beta-p^\beta v^\alpha +\frac{gq}{2m}
\Bigr( -\frac{p\cdot v}{m}\Bigr) 
\Bigl(S^{\mu\alpha}F_\mu^{~\beta}- S^{\mu\beta}F_\mu^{~\alpha}\Bigr),
\label{eq:basic2}
\end{eqnarray}
%
%==========================%
where $\mu=0,1,2,3$ and 
$v^\mu=dx^\mu(\tau)/d\tau$. $p^\mu$ and $S^{\mu\nu}$ are the 
4-momentum and the angular momentum tensor, respectively. $g$ is the 
gyromagnetic ratio of the probe particle. ``$ ~\cdot~ $'' denotes the 
inner product with respect to the metric, that is, $ X\cdot Y =g_{\mu\nu}
X^\mu  Y^\nu $. We adopt the normalisation $v \cdot v =-1$. 
As we want to set the orbit 
$x^\mu(\tau)$ to be the center of the mass, we impose 
the ``supplementary condition'' such that\cite{Wald} 
%========<Equation>========%
%
\begin{eqnarray}
p^\mu S_{\mu \nu}=0. \label{eq:centre}
\end{eqnarray}
%
%==========================%
It is well known in general that 
$p^\mu$ is not parallel to $v^\mu$ due to the spin. 
From Eqs. (\ref{eq:basic1}) and (\ref{eq:basic2}) with the condition 
(\ref{eq:centre}), we obtain the relation between $v^\mu$ and $p^\mu$,
%========<Equation>========%
%
\begin{eqnarray}
v^\mu \simeq \frac{1}{mf}\Bigl[ p^\mu -\frac{q}{m^2f^2} 
S^{\mu\nu}F_{\nu\alpha}p^\alpha 
\Bigl(1-\frac{g}{2} f\Bigr) \Bigr] \label{eq:VP}
\end{eqnarray}
%
%==========================%
in the lowest order needed for the later evaluation of 
the gravitational spin-spin interaction and the slow rotating 
limit of the background space-time. In Eq. (\ref{eq:VP}), the
function, $f$, is defined by 
%========<Equation>========%
%
\begin{eqnarray}
f=\frac{1}{m}\frac{p \cdot p}{v \cdot p}. \label{eq:func}
\end{eqnarray}
%
%==========================%
The exact expression of Eq. (\ref{eq:VP}) is given in appendix A. 
Due to the spin and electromagnetic field term, 
the norm of the momentum is not conserved along the orbit of the 
probe particle\cite{Foot1}, 
%========<Equation>========%
%
\begin{eqnarray}
v^\nu \nabla_\nu (p \cdot p) = \frac{gq}{2m}
\Bigl(\frac{p \cdot v}{-m} \Bigr)p^\mu
\nabla_\mu F_{\alpha \beta} S^{\alpha \beta}. \label{eq:PP}
\end{eqnarray}
%
%==========================%
On the other hand $S^{\mu\nu}S_{\mu\nu}$ is exactly conserved, 
%========<Equation>========%
%
\begin{eqnarray}
v^\mu \nabla_\mu (S^{\alpha \beta}S_{\alpha \beta})=0. \label{eq:SS}
\end{eqnarray}
%
%==========================%

The quantity $Q_\kappa$ defined by 
%========<Equation>========%
%
\begin{eqnarray}
Q_\kappa=(p^\mu +q A^\mu)\kappa_\mu +\frac{1}{2}S^{\mu \nu}\nabla_\mu
\kappa_\nu, 
\end{eqnarray}
%
%==========================%
is a conserved if $\kappa$ is a Killing vector and satisfies 
$\mbox{\pounds}_\kappa A^\mu=0 $ and 
$\mbox{\pounds}_\kappa F_{\mu \nu}=0$ since 
%========<Equation>========%
%
\begin{eqnarray}
v^\mu \nabla_\mu Q_\kappa = qv_\mu \mbox{\pounds}_\kappa A^\mu 
+\frac{gq}{4m} \Bigl(\frac{p \cdot v}{-m}\Bigr)
S^{\mu \nu} \mbox{\pounds}_\kappa F_{\mu \nu}.
\end{eqnarray}
%
%==========================%

\section{A Probe Particle on Kerr-Newman-deSitter Black Hole and Force
Balance}

In this section, 
we confirm the force cancellation including the gravitational spin-spin 
interaction between a probe particle and the central 
Kerr-Newman-deSitter(KNdS) black hole. For this purpose we  omit the 
higher order term of the angular momentum. Namely, we take 
the slow rotating limit and the metric becomes 
%========<Equation>========%
%
\begin{eqnarray}
ds^2 = -V(r)dt^2+\frac{1}{V(r)}dr^2+r^2 d\Omega^2_2+2a(V-1){\rm sin}^2
\theta dt d\varphi+O(a^2),  \label{eq:metric}
\end{eqnarray}
%
%==========================%
where $a$ is the parameter of the angular momentum and 
$V(r)=1-2M/r+Q^2/r^2-(\Lambda/3)r^2$. $\Lambda=3H^2$ is a positive 
cosmological constant. The exact expression is given in the 
appendix A. The vector potential is 
%========<Equation>========%
%
\begin{eqnarray}
A = -\frac{Q}{r}(dt -a{\rm sin}^2\theta d\varphi)
+O(a^2).\label{eq:pot}
\end{eqnarray}
%
%==========================%
This system has two Killing vectors $\xi=\partial_t$ and $\psi = 
\partial_\varphi$. These Killing vectors satisfy $\mbox{\pounds}_\kappa
A^\mu=0$ and $\mbox{\pounds}_\kappa F_{\mu\nu}=0$ and thus we have two 
constants of motion.

The constants of the motion related to the Killing vectors, $\xi$ and 
$\psi$,  become 
%========<Equation>========%
%
\begin{eqnarray}
-Q_\xi & = & -\xi^\mu  (p_\mu+qA_\mu)-\frac{1}{2}S^{\mu\nu}
\nabla_\mu \xi_\nu\nonumber \\
& = & Vp^0-a(V-1){\rm sin}^2\theta
p^\varphi+\frac{qQ}{r}-\frac{aS}{r^2}(V-1){\rm cos}^2\theta
\end{eqnarray}
%
%==========================%
and
%========<Equation>========%
%
\begin{eqnarray}
Q_\psi & = & \psi^\mu  (p_\mu+qA_\mu)+\frac{1}{2}S^{\mu\nu}
\nabla_\mu \psi_\nu \nonumber \\
& = & a(V-1){\rm sin}^2\theta p^0+r^2{\rm sin}^2\theta p^\varphi
+\frac{aqQ}{r}{\rm sin}^2\theta+S{\rm cos}^2\theta, 
\end{eqnarray}
%
%==========================%
respectively. Now we assume $v^\theta=p^\theta=0$. 
Setting $Q_\xi=-m$ and $Q_\psi=S {\rm cos}^2\theta$, 
$p^0$ and $p^\varphi$ can be written as 
%========<Equation>========%
%
\begin{eqnarray}
p^0=\frac{m}{V}\Bigl(1-\frac{qQ}{mr} \Bigr)~~~~{\rm and}~~~~
p^\varphi=-\frac{aqQ}{r^3} -\frac{ma(V-1)}{r^2V}\Bigl(1-\frac{qQ}{mr}
\Bigr).
\end{eqnarray}
%
%==========================%
%or $v^0$ and $v^\varphi$ are 
%========<Equation>========%
%
%\begin{eqnarray}
%v^0 \simeq \frac{1}{fV}\Bigl(1-\frac{qQ}{mr} \Bigr)~~~~{\rm and}~~~~
%v^\varphi \simeq 
%-\frac{aqQ}{mr^3} -\frac{a(V-1)}{Vr^2} \Bigl(1-\frac{qQ}{mr} \Bigr).
%\end{eqnarray}
%
%==========================%

For simplicity we also assume that the spin vector ${\vec {S}}$ 
of the probe particle is parallel to that of the background 
space-time, ${\vec {S}}=(0,0,S)$ in a sort of Cartesian coordinate. In the 
present coordinate of Eq. (\ref{eq:metric}), each components of the 
spin tensor $S^{\mu\nu}$ are written as  
%========<Equation>========%
%
\begin{eqnarray}
S^{\theta\varphi}=\frac{{\rm cos}\theta}{{\rm sin}\theta}\frac{S}{r^2}
\nonumber 
\end{eqnarray}
%
%==========================%
and $S^{r\theta}=S^{r\varphi}=0$. In this case, Eq. (\ref{eq:PP}) becomes 
%========<Equation>========%
%
\begin{eqnarray}
v^\mu \nabla_\mu (p \cdot p) =\frac{gq}{m}\Bigl( \frac{p \cdot v}{-m}\Bigl)
p^\mu \nabla_\mu F_{\theta\varphi}S^{\theta\varphi}+O(aS^2) 
\label{eq:VVPP}
\end{eqnarray}
%
%==========================%
and Eq. (\ref{eq:SS}) concludes that $S$ is a constant. 

Let us determine the reading order of the function, $f$, defined by 
Eq. (\ref{eq:func}). First of all, Eq. (\ref{eq:VP}) yields
%========<Equation>========%
%
\begin{eqnarray}
& & v^0=\frac{1}{mf}p^0,~~v^r=\frac{1}{mf}p^r,~~v^\theta
=\frac{1}{mf}p^\theta+O(a^2), ~~v^\varphi=\frac{1}{mf}
p^{\varphi}+O(a^2S) 
\label{eq:speed}
\end{eqnarray}
%
%==========================%
which implies 
%========<Equation>========%
%
\begin{eqnarray}
p^\mu p_\mu = (mf)^2 v_\mu v^\mu+O(a^2)=-(mf)^2+O(a^2).
\end{eqnarray}
%
%==========================%
Substituting these approximate expression into Eq. (\ref{eq:VVPP}), we 
obtain the equation for the function $f$:
%========<Equation>========%
%
\begin{eqnarray}
v^\mu \partial_\mu {\rm ln}f^2= -\frac{gaSqQ}{m^2} v^\mu
\partial_\mu \Bigl(\frac{1}{r^3}\Bigr) {\rm cos}^2\theta+O(a^2),
\end{eqnarray}
%
%==========================%
and then 
%========<Equation>========%
%
\begin{eqnarray}
f & \simeq & e^{-\frac{gaSqQ}{m^2r^3}{\rm cos}^2\theta} \nonumber \\
& \simeq & 1- \frac{gaSqQ}{m^2r^3}{\rm sin}\theta {\rm cos}\theta
=:1-g{\tilde f},\label{eq:fufufu} 
\end{eqnarray}
%
%==========================%
where ${\tilde f}=\frac{aSqQ}{m^2r^3}{\rm cos}^2 \theta$.

Now we can calculate the force acting on the probe particle. Noting 
%========<Equation>========%
%
\begin{eqnarray}
v^\mu \partial_\mu p^r & = & mv^\mu \partial_\mu (fv^r) \nonumber \\
& = & mf v^\mu \partial_\mu v^r+m(v^r)^2\partial_r f, 
\end{eqnarray}
%
%==========================%
we see that the equation of motion becomes 
%========<Equation>========%
%
\begin{eqnarray}
mf v^\mu \partial_\mu v^r = -m(v^r)^2\partial_r f 
-\Gamma^r_{\alpha\beta}v^\alpha p^\beta -
\frac{1}{2}R^r_{~\nu \alpha\beta}v^\nu S^{\alpha\beta}
+qF^r_{~\nu}v^\nu+\frac{gq}{4m}\Bigr( -\frac{p\cdot v}{m}\Bigr) 
\nabla^r F_{\alpha\beta}S^{\alpha\beta}=:m fF^r,
\end{eqnarray}
%
%==========================%
where $F^\mu$ denotes the force per unit mass. Using Eqs. 
(\ref{eq:speed}) $\sim$ (\ref{eq:fufufu}), each terms in the 
right-hand side are evaluated as follows,
%========<Equation>========%
%
\begin{eqnarray}
& & -m(v^r)^2\partial_r f   \simeq  gm(1-V)\partial_r {\tilde f} 
\simeq -\frac{gaSqQ}{r^2m}\Lambda {\rm cos}^2 \theta \label{eq:part1}\\ 
& & -f^{-1}\Gamma^r_{\mu\nu} v^\mu p^\nu \simeq -\frac{m}{2}V'
=-\frac{Mm}{r^2}+\frac{mQ^2}{r^3}+m \frac{1}{3}\Lambda r \\
& & f^{-1}qF^r_{~\mu}v^\mu \simeq f^{-1}qVA_0'v^0 \simeq \frac{qQ}{r^2}
-\frac{q^2Q^2}{mr^3} \\
& & 
-\frac{1}{2}f^{-1}R^r_{~\mu\alpha\beta} v^\mu S^{\alpha\beta}
\simeq -R^r_{~0\theta\varphi} v^0 S^{\theta\varphi} = \frac{6aSM}{r^4}
{\rm cos}^2\theta \\ 
& & 
\frac{gq}{4m}f^{-1}\Bigr( \frac{p\cdot v}{-m}\Bigr) 
\nabla^r F_{\alpha\beta}S^{\alpha\beta} \simeq  
\frac{gq}{2m}\nabla^r F_{\theta\varphi} S^{\theta\varphi}
\simeq -V\frac{3gaSqQ}{mr^4}{\rm cos}^2\theta 
\simeq  -\frac{3gaSqQ}{mr^4}{\rm cos}^2\theta+\frac{gaSqQ}{r^2m}\Lambda 
{\rm cos}^2\theta \label{eq:part2}
\end{eqnarray}
%
%==========================%
Summing up Eqs. (\ref{eq:part1})$\sim$(\ref{eq:part2}), 
we obtain the total force,  
%========<Equation>========%
%
\begin{eqnarray}
F^r =\frac{d^2r}{d\tau^2} \simeq \frac{1}{3}\Lambda r  -\frac{Mm-Qq}{mr^2}
+\frac{(m^2-q^2)Q^2}{m^2 r^3}+
\frac{3aS(2Mm-gQq)}{m^2r^4}{\rm cos}^2\theta.  \label{eq:force}
\end{eqnarray}
%
%==========================%
Eq (\ref{eq:force}) obviously yields 
$F^r \simeq \frac{1}{3}\Lambda r=H^2r$ as $Q=M, q=m$ and $g=2$. 
This expression of the force is same as pure deSitter space-time case.  
That is, the force cancellation holds including the order of the 
gravitational spin-spin interaction except for the repulsive 
force due to the cosmological constant. We can show that $F^\theta$ 
and $F^\varphi$ are negligible under the approximation used here. 
 
In the asymptotically deSitter space-times, it is, however, 
not trivial to estimate the gyromagnetic ratio, $g_{\rm BG}$, of a background
space-time because the metric component $g_{t\varphi}$, which is
related to the angular momentum, contains 
the cosmological constant and there is not the 
adequate definition of the total angular momentum $J$. If $J$ is 
physically defined, the gyromagnetic ratio will be given by 
$g_{\rm BG}=2Ma/J$. Subtracting 
the cosmological constant term from the definition of the total 
angular momentum,  
$(1/16\pi)\int_{S_\infty}dS_{\mu\nu}\nabla^\mu \psi^\nu$\cite{Wald2},
we are resulted in $J=Ma$ and $g_{\rm BG}=2$.

\section{Summary and Discussion}

In this paper we confirmed the force cancellation including the gravitational 
spin-spin interaction between a probe particle and the 
slow rotating Kerr-Newman-deSitter(KNdS) space-time.  We remind
readers that the KNdS space-time with $Q=M$ 
is free from naked singularities in the slow rotating limit. 
The force balance holds when $Q=M$, $q=m$ and $g=2$. 
As we said, this fact may suggest the existence of the exact solution which 
is the spinning version of the Kastor-Traschen solution. 
It is important to remind you that one of the present authors 
found a new exact solution which is the spinning dilatonic version 
of the Kastor-Traschen solution in the Einstein-Maxwell-dilaton-axion 
theory with a positive cosmological constant\cite{Tess3}.

Even in asymptotically deSitter space-times, the force balance 
indicates the existence of a symmetry which might be slightly 
related to supersymmetry. The symmetry should give the relation 
between components of the metric and the electromagnetic field and 
help us to solve the Einstein equation. More precisely, the energy 
bound as the Bogomol'nyi bound holds and a symmetry appears when 
the bound is saturated. The saturation yields the Killing spinor 
like equation related to supersymmetry and the equation gives a 
relation between components. In fact, we can easily show the 
bound nature in  certain case without magnetic field as it is shown 
in the appendix B. 

The definition of the total angular momentum is also important in 
asymptotically deSitter space-times and should be investigated 
although it is not directly related to our main aim. 

\vskip1cm

\centerline{\bf Acknowledgment}
We  would like to thank Katsuhiko Sato, Misao Sasaki and 
Yasushi Suto for their 
encouragements. TS thanks Gary W. Gibbons and DAMTP 
relativity group for their hospitality at Cambridge. TS also thanks 
Takahiro Tanaka for his kind response at the beginning of this study. 
This work is partially supported by the JSPS[No.310(TS)]. 

\appendix

\section{Some Exact Expressions}

The relation between $v^\mu$ and $p^\mu$ is derived by 
$v^\alpha \nabla_\alpha (p^\mu S_{\mu\nu})=0$. After some tedious 
calculation, we obtain 
%========<Equation>========%
%
\begin{eqnarray}
v^\mu=\frac{p \cdot v}{p \cdot p}
\Bigl[p^\mu+\frac{1}{(-p\cdot p)}
\frac{\frac{1}{2}R_{\nu\rho\alpha\beta}
S^{\alpha\beta}p^\rho S^{\mu\nu}-q\Bigl( 1-\frac{g}{2m} \frac{p\cdot
p}{p \cdot v}\Bigr)F_{\alpha\beta}p^\beta S^{\mu\alpha}-gq
\Bigl( \frac{p \cdot v}{-m}\Bigr)\nabla_\nu
F_{\alpha\beta}S^{\alpha\beta}
S^{\mu\nu}  }{1+\frac{1}{(-p \cdot p)} \Bigl( R_{\mu\nu\alpha\beta}S^{\mu\nu}
S^{\alpha\beta}-\frac{q}{2}F_{\mu\nu}S^{\mu\nu}\Bigr)  }\Bigr].
\end{eqnarray}
%
%==========================%
and yields Eq. (\ref{eq:VP}) under an appropriate approximation. 

Next, we describes the exact expression of the KNdS space-time
\cite{Carter}\cite{Moss}. The metric and the vector potential are 
%========<Equation>========%
%
\begin{eqnarray}
ds^2=\frac{\rho^2}{\Delta}dr^2
+\frac{\rho^2}{1+\frac{\Lambda}{3}a^2{\rm cos}^2\theta } d\theta^2
+\Bigl( 1+\frac{\Lambda}{3}a^2{\rm cos}^2\theta   \Bigr)  
\frac{a^2 {\rm sin}^2\theta}{\chi^4\rho^2}
\Bigl(dt-\frac{\sigma^2}{a}d \phi \Bigr)^2 
-\frac{\Delta}{\chi^4 \rho^2}(dt-a{\rm sin}^2\theta d \phi)^2
\end{eqnarray}
%
%==========================%
and
%========<Equation>========%
%
\begin{eqnarray}
A=-\frac{Qr}{\chi^2\rho^2}(dt - a {\rm sin}^2 \theta d \phi),
\end{eqnarray}
%
%==========================%
where
%========<Equation>========%
%
\begin{eqnarray}
& & \sigma^2=a^2+r^2 \nonumber \\
& & \Delta=(a^2+r^2)\Bigl(1-\frac{\Lambda}{3}r^2 \Bigr) -2Mr+Q^2
\nonumber \\
& & \rho^2=r^2+a^2{\rm cos}^2 \theta \nonumber \\
& & \chi^2 = 1+\frac{\Lambda}{3}a^2. \nonumber 
\end{eqnarray}
%
%==========================%
Up to the order of $O(a)$ we obtain Eqs. (\ref{eq:metric}) and 
(\ref{eq:pot}).

\section{Energy Bound Theorem in a certain case}

This appendix is based on  the
Refs. \cite{Bogo1}\cite{Bogo2}\cite{Bogo3}. For simplicity, we 
consider the case in which the magnetic component is absent. However, 
we guess that the extension into the case with the magnetic field as
well as the electric field is easy task. We define the derivative 
operator on a spinor $\epsilon$ as 
%========<Equation>========%
%
\begin{eqnarray}
{\hat {\nabla}}_i \epsilon = \Bigl ( D_i+\frac{1}{2}K_{ij}\gamma^j 
\gamma^{\hat 0}+\frac{i}{2}H\gamma_i -\frac{i}{2}\gamma^{\hat
0}\gamma^j \gamma_i E_j \Bigr)\epsilon, \label{eq:derivative}
\end{eqnarray}
%
%==========================%
where $i=1,2,3$, $D_i\epsilon =(\partial_i+{}^{(3)}\Gamma_i) \epsilon$ and 
${}^{(3)}\Gamma_i=-(1/8)e^{j{\hat k}} D_i e^{\hat \ell}_j
[\gamma_{\hat \ell}, \gamma_{\hat k}]$. $E^i$ is the electric field 
vector. $e^{\hat i}_j$, $D_i$ and $K_{ij}$ are a unit-orthogonal basis,
a covariant derivative and the extrinsic curvature 
of three dimensional spacelike
hypersurfaces, respectively. 
The expression of Eq. (\ref{eq:derivative}) 
is inspired by $N=2$ supergravity with a {\it 
negative} cosmological constant. Although the derivative operator of
Eq. (\ref{eq:derivative}) is not covariant form in the full space-time, 
it is enough to argue the energy bound in the present situation. 
$\epsilon$ is assumed to satisfy the 
modified Witten equation 
%========<Equation>========%
%
\begin{eqnarray}
\gamma^i {\hat \nabla}_i \epsilon =0.\label{eq:witten}
\end{eqnarray}
%
%==========================%

In the same way as \cite{Bogo1}\cite{Bogo2}\cite{Bogo3}, 
we obtain the identity
%========<Equation>========%
%
\begin{eqnarray}
D^i(\epsilon^\dagger {\hat \nabla}_i \epsilon)
& = & ({\hat \nabla}_i\epsilon )^\dagger ({\hat \nabla}^i\epsilon)+
\frac{1}{4}\epsilon^\dagger \Bigl[{}^{(3)}R+K^2-K_{ij}K^{ij}-6H^2  
\Bigr] \epsilon 
+\frac{1}{2}\epsilon^\dagger D_i(K^i_j-\delta^i_j K)\gamma^j
\gamma^{\hat 0} \epsilon \nonumber \\
& & ~~~-\frac{1}{2}\epsilon^\dagger E^i E_i \epsilon 
-i \epsilon^\dagger D^iE_i \gamma^{\hat 0}\epsilon. \label{eq:before}
\end{eqnarray}
%
%==========================%
Taking the volume integral of the above 
and using a parts of the Einstein and Maxwell equations, 
%========<Equation>========%
%
\begin{eqnarray}
\int_{S_\infty} dS_i \epsilon^\dagger {\hat \nabla}_i \epsilon
= \int_V dV \Bigl[ ({\hat \nabla}_i\epsilon )^\dagger ({\hat \nabla}^i 
\epsilon)+  4 \pi \epsilon^\dagger (T_{{\hat 0}{\hat 0}}+T_{{\hat 0}{\hat
i}}\gamma^{\hat i}\gamma^{\hat 0}) \epsilon+4 \pi i \epsilon^\dagger \rho_e 
\gamma^{\hat 0}\epsilon \Bigr], \label{eq:intel}
\end{eqnarray}
%
%==========================%
where $T_{\mu\nu}$ is the energy momentum tensor
subtracted by the energy of the electric field and 
$\rho_e$ is the electric charge density.  
The left-hand side of Eq. (\ref{eq:intel}) can be written as
%========<Equation>========%
%
\begin{eqnarray}
\int_{S_\infty} dS^i \epsilon^\dagger {\hat \nabla}_i \epsilon
=\int_{S_\infty} dS^i \epsilon_0 (\Gamma_i'-\gamma_i \gamma^j
\Gamma_j')\epsilon_0, \label{eq:surface}
\end{eqnarray}
%
%==========================%
where 
%========<Equation>========%
%
\begin{eqnarray}
\Gamma_i'={}^{(3)}\Gamma_i+\frac{1}{2}H\gamma_i-\frac{i}{2}\gamma^{\hat 0}
\gamma^j \gamma_i E_j \label{eq:gamma}
\end{eqnarray}
%
%==========================%
and $\epsilon_0$ is a constant spinor satisfying $\gamma^{\hat
 0} \epsilon_0 =-i \epsilon_0$. 
Inserting Eq. (\ref{eq:gamma}) into Eq. (\ref{eq:surface}), we obtain 
%========<Equation>========%
%
\begin{eqnarray}
\int_{S_\infty} dS^i \epsilon^\dagger {\hat \nabla}_i \epsilon
=\frac{1}{4}\int_{S_\infty} dS^i \epsilon_0 (\partial_j h_i^{~j}-\partial_i 
h^j_{~j})\epsilon_0
+\frac{1}{2}\int_{S_\infty} dS_i  \epsilon_0^\dagger (K^i_{~j}-\delta^i_j
K+2H\delta^i_j)\gamma^j \gamma^{\hat 0} \epsilon_0
-\int_{S_\infty} dS_i \epsilon_0^\dagger E^i \epsilon_0, \label{eq:final}
\end{eqnarray}
%
%==========================%
where $h_{\mu\nu}=g_{\mu\nu}-{\overline g}_{\mu\nu}$ and ${\overline g}
_{\mu\nu}$ is the metric of the deSitter space-time. 
The first and second term in the right-hand side 
are exactly the ADM energy and the net 3-momentum. 
The last term is the total electric charge. 
Note that we have to take the asymptotic region
carefully.  The situation differs from asymptotic flat case in which  
the extrinsic curvature of the spacelike hypersurface behaves 
$K^i_j \to 0 $ towards the spatial infinity. In asymptotically
deSitter space-times, 
the slices with $K \to 3H $ as $r \to \infty $
are best to define the conserved charges\cite{Bogo3}. Moreover, 
we note that the 
existence of the solution for the modified Witten equation is guaranteed 
and the surface integral $\int_{S_\infty} dS_i 
\epsilon^\dagger {\hat \nabla}^i \epsilon $ can be written as Eq. 
(\ref{eq:surface}) 
if the traceless part of the extrinsic curvature has the behaviour  
${\tilde K}^i_j=O(1/r^2)$ near the infinity. This implies that the
second term in the right-hand side of Eq. (\ref{eq:final}) 
vanishes\cite{Bogo3}. 

Finally, Eqs. (\ref{eq:intel}) and
(\ref{eq:final}) lead us the result of the inequality $M_{\rm ADM} \geq
|Q|$ under the dominant energy condition and the causality condition of the
electric current. The saturation yields ${\hat \nabla}_i \epsilon =0$ 
slightly related to $N=2$ supersymmetry. As the Majumdar-Papapetrou
solution\cite{Hull}, the Kastor-Traschen solution has the super covariantly
constant spinor $\epsilon_0$ satisfying 
$\gamma^{\hat 0}\epsilon_0=-i \epsilon_0$. This also strongly 
suggests the existence of multi-spinning black-hole solution in 
asymptotically deSitter space-times if we can probe the general 
energy bound argument including the magnetic field.

%======================================%
%<<<<<<<<<<<< REFERENCES >>>>>>>>>>>>>>%
%======================================%

%\endmulticols

\begin{thebibliography}{22}
\bibitem{KT}
D. Kastor and J. Traschen, Phys. Rev. {\bf D47}, 5370(1993)
\bibitem{Brill} 
D. R. Brill, G. T. Horowitz, D. Kastor and J. Traschen, 
Phys. Rev. {\bf D49}, 840(1994) 
\bibitem{Nakao}
K. Nakao, T. Shiromizu and S. A. Hayward, Phys. Rev. {\bf D52}, 796(1995);\\
D. Ida, K. Nakao, M. Siino and S. A. Hayward, Phys. Rev. 
{\bf D58}, 121501(1998) 
\bibitem{nohair}
G. W. Gibbons and S. W. Hawking, Phys. Rev. {\bf D15}, 2738(1977);\\
S. W. Hawking and I. G. Moss, Phys. Lett. {\bf 110B}, 35(1982);\\
T. Shiromizu, K. Nakao, K. Maeda and H. Kodama, Phys. Rev. {\bf D47}, 
R3099(1993);\\
S. A. Hayward, T. Shiromizu and K. Nakao, Phys. Rev. {\bf D49},
5080(1994);\\
K. Maeda, T. Koike, M. Narita and A. Ishibashi, Phys. Rev. {\bf D57}, 
3503(1998) 
\bibitem{Bad}
D. L. Welch, Phys. Rev. {\bf D52}, 985(1995) 
\bibitem{MP}
S. D. Majumdar, Phys. Rev. {\bf 72}, 930(1947);\\
A. Papapetrou, Proc. Roy. Irish Acad. {\bf A51}, 191(1947)
\bibitem{IW}
W. Israel and G. A. Wilson, J. Math. Phys. {\bf 13}, 865(1972)
\bibitem{Perjes}
Z. Perjes, Phys. Rev. Lett. {\bf 27}, 1668(1971)
\bibitem{HH}
J. B. Hartle and S. W. Hawking, Commun. Math. Phys. {\bf 26}, 87(1972)
\bibitem{Sen1}
A. Sen, Nucl. Phys. {\bf B440}, 421(1995) 
\bibitem{Sen2}
G. T. Horowitz and A. Sen, Phys. Rev. {\bf D53}, 808(1996)
\bibitem{KT2}
J. Tiomno, Phys. Rev. {\bf D7}, 35(1973);\\
W. Israel and J. T. J. Spanos, Lett. Nuovo Cim. {\bf 7}, 245(1973);\\
D. Kastor and J. Traschen, Class. Quantum Grav. {\bf 16}, 1265(1999)
\bibitem{Papa}
A. Papapetrou, Proc. Roy. Soc. Lond. {\bf A209}, 248(1951)
\bibitem{Dixon}
W. G. Dixon, Proc. Roy. Soc. Lond. {\bf A314}, 499(1970)
\bibitem{Wald}
R. M. Wald, Phys. Rev. {\bf D6}, 406(1972) 
\bibitem{Tess1}
T. Shiromizu, Phys. Rev. {\bf D60}, 104046(1999) 
\bibitem{EMDA}
R. Kallosh, D. Kastor, T. Ortin and T. Torma, Phys. Rev. {\bf D50}, 
6374(1994);\\
E. Bergshoeff, R. Kallosh and T. Ortin, Nucl. Phys. {\bf 478}, 156(1996)
\bibitem{Tess2}
T. Shiromizu, Phys. Lett. {\bf B460}, 141(1999)
\bibitem{spin1}
A. R. Prasanna and K. S. Virbhadra, Phys. Lett. {\bf A138},
242(1989);\\
K. S. Virbhadra and A. R. Prasanna, Phys. Lett. {\bf A145}, 410(1990)
\bibitem{Foot1}
In the case with $g=2$, Eqs. (\ref{eq:basic1}) and (\ref{eq:basic2}) 
can be derived from the equation without the 5-dimensional gauge field 
in five dimensions via the usual Kaluza-Klein reduction. In the five 
dimensions, the norm of the 5-momentum, $p^M p_M$, is conserved along the 
orbit of the probe particle, where  $M=0,1,2,3,4$. Obviously, 
$v^N \nabla_N (p^M p_M)=0$ should become Eq. (\ref{eq:PP}) in the 
terms of the four dimensions. 
\bibitem{Wald2}
For example, R. M. Wald, Problem 6 in Chapter 11, {\it General Relativity}
(The University of Chicago Press, Chicago 1984)
%\bibitem{Ernst}
%F. J. Ernst, Phys. Rev. {\bf 167}, 1175(1968)
\bibitem{Carter}
B. Carter, in Black Holes, ed. V. De Sabbata and Z. Zhang(Amsterdam:
Kluwer Academic Publishes, 1983)
\bibitem{Moss}
C. M. Chambers and I. G. Moss, Class. Quantum Grav. {\bf 11}, 1035(1994)
\bibitem{Bogo1}
G. W. Gibbons, S. W. Hawking, G. T. Horowitz and M. J. Perry, Commun. Math. 
Phys. {\bf 88}, 295(1983)
\bibitem{Bogo2}
D. Kastor and J. Traschen, Class. Quantum Grav. {\bf 13}, 2753(1996)
\bibitem{Bogo3}
T. Shiromizu, Phys. Rev. {\bf D60}, 064019(1999) 
\bibitem{Tess3}
T. Shiromizu, to be published in Prog. Theor. Phys, hep-th/9910176 
\bibitem{Hull}
G. W. Gibbons and C. M. Hull, Phys. Lett. {\bf 109B},190(1982) 
\end{thebibliography}
\end{document}